\documentstyle[12pt,aaspp4] {article}

\def\etal{et al.\rm}
\def\f555w{{\it F555W}\/}

\righthead{Q0957 Precision Photometry}

\begin{document}

\title{Precision Photometry for Q0957+561 Images A and B}

\author{Wesley N. Colley and Rudolph E. Schild}
\affil{Harvard-Smithsonian Center for Astrophysics, 60 Garden Street, Cambridge
MA 02138}

\begin{abstract}

Since the persuasive determination of the time-delay in Q0957+561, much
interest has centered around shifting and subtracting the A and B light-curves
to look for residuals due to microlensing.  Solar mass objects in the lens
galaxy produce variations on timescales of decades, with amplitudes of a few
tenths of a magnitude, but MACHO's (with masses of order $10^{-3}$ to
$10^{-7}M_\odot$) produce variations at only the 5\% level.  To detect such
small variations, highly precise photometry is required.

To that end, we have used 200 observations over three nights to examine the
effects of seeing on the light-curves.  We have determined that seeing itself
can be responsible for correlated 5\% variations in the light-curves of A and
B.  We have found, however, that these effects can be accurately removed, by
subtracting the light from the lens galaxy, and by correcting for cross
contamination of light between the closely juxtaposed A and B images.  We find
that these corrections improve the variations due to seeing from 5\% to a level
only marginally detectable over photon shot noise (0.5\%).

\end{abstract}


\section{Introduction}

The attempts to observe quasar brightness fluctuations to measure time delays
and detect microlensing have shown that a high level of photometric accuracy is
desirable. For Q0957+561 (Walsh \etal\ 1979) a substantial controversy about
the time delay persisted for more than a decade despite reliable photometry on
1000 nights (Kundi\'c et al. 1997, Schild 1990, Vanderriest \etal\ 1989, Press
\etal\ 1992b, Haarsma \etal\ 1996).  Part of the confusion could have been due
to microlensing, which is difficult to distinguish from intrinsic quasar
fluctuations, upon which the time delay measurements are leveraged.  The
observational evidence suggests that microlensing has a component due to solar
mass stars causing fluctuations on time scales of decades and amplitudes of a
few tenths of a magnitude (Schild \& Smith 1991, Pelt \etal\ 1998). However, it
is possible that there exists a more rapid microlensing component due to
MACHO's in the lens galaxy's halo (Schild 1996, Young 1981, Gott 1981).  Such
fluctuations occur at the few percent level; thus photometry of both images is
required to have a precision of 0.01 magnitudes or better.

A very large data base for the Q0957 system has been collected from CCD data
frames since 1979 and reduced on several generations of computers, but the
basic photometric procedure has been the same.  Aperture photometry with
$6\arcsec$ diameter apertures, and correction for an $R \approx 18.3$ lens
galaxy in the B aperture, have been the basis throughout, and many procedural
details and error considerations have been given in the appendix of Schild and
Cholfin 1986. The most problematic aspect of the reduction scheme is the
bleeding of light from one quasar image to the aperture of the other,
particularly when seeing is poor due to turbulence in the terrestrial
atmosphere. In principle it should be possible to estimate and correct for the
effect by examining the spread of light from nearby field stars, but attempts
to implement correction schemes have failed in the past, probably because they
have not taken into account the light of the lens galaxy G1, which has an
extended profile and is situated between the two quasar images. Thus it has
been recognized that significant improvement in the photometric precision would
be achieved only when an image from the Hubble Space Telescope could be
analyzed to measure the brightness profile of lens galaxy G1.

Bernstein \etal\ (1997) have recently used the HST to image the lens system, so
that it is now possible to make accurate estimates of the contribution of light
from lens galaxy G1 to the A and B apertures, and thence, to correct for the
spillover of light from one quasar image into the aperture of the other. The
purpose of this contribution is to establish the method of improved aperture
photometry, and to demonstrate its use on a data set consisting of 200 images
collected on three nights in February, 1995, when both quasar images were not
showing large brightness fluctuations. Of course the most trustworthy
observation of quasar variability is one which shows no variability, and the
available data allow us to effectively compare our measurement uncertainty
limits with Poisson statistics.

\section{Photometry Scheme}

For over a decade, RS has monitored Q0957 and amassed a ``master dataset'' of
the variations of images A and B (Schild \& Thompson 1995).  Our 1k$\times$1k
CCD images have pixel scales of one-third of an arcsecond, exposure times of
450 seconds, read-noise of $8e^-$, and FWHM seeing of 1.5--2 arcseconds.  This
generally allows for about 5 unsaturated, well sampled comparison stars in the
$R$-band on each data-frame.  There are thousands of such frames, and together,
they represent one of the largest continual monitoring efforts in history.

Since the amount of archived data is large, an automated photometry code (one
that does not require mouse clicks, or manual entry of parameters) is
desirable.  Interfacing all of the necessary photometry, pre-processing and
post-processing code proved to be easiest for WNC in the IDL programming
language, at cost of some redundancy with available photometry packages in
IRAF.

The photometry algorithms we have used are modelled after those in IRAF, {\it
daofind}, for example.  First the flat-fielded image is convolved with a
``lowered Gaussian'' kernel with width similar to that of the seeing.  The
lowered Gaussian is a two-dimensional Gaussian bell, less a constant.  The
constant equals the volume under the Gaussian, out to two-sigma, where the
convolution kernel terminates.  A threshold is selected in terms of the
standard deviation of noise in the sky, and pixels above that threshold are
tabulated.  We then use a friend-of-friend algorithm to group adjacent high
pixels from the list.  These groups are presumed to reside at the locations of
sources.  Typically, this algorithm locates about 15--20 sources per frame.

In order to understand which source is which, we first generate a template list
of sources, which contains only the 10 sources we are interested in, two of
which are the QSO images A and B.  This list contains only the relative
astrometry for each source (in pixels).  For each data frame, we generate
aperture astrometry for each of the sources located by the friend-of-friend
algorithm above.  We then compute the pixel position angles of each source
relative to a particular source in the template and data source list.  The
correlation function of the data angles with the template angles reveals a peak
at the correct angular offset between data sources and template sources.  One
can iterate for a few sources, and determine the magnification, rotation and
translation of the data source list relative to the template source list, from
which the positions of all the template stars on the data frame are determined.
This scheme is particularly useful, because without clever thresholding the
quasar images tend to blend together, or disappear completely.  But after the
geometry of the field is determined, one can compute the locations of the
quasar images with little difficulty.

The final complication is cosmic rays.  Since we would like to use each
exposure, we do not have the luxury of medianing several exposures to remove
cosmic rays.  Furthermore, we would like to avoid visual inspection as the
cosmic ray detector.  Our automated cosmic ray detection scheme generates a
model point spread function by co-adding several of the sources on each frame
(we will need this model p.s.f. for the galaxy subtraction anyway).  We then
fit this model to each of the sources.  A large spike in the post-fit residual
from any source is interpreted as a cosmic ray, and the source is thrown out of
the reduction.  Only occasionally (a few times per night) is a source lost from
the reduction.

With the useful source positions now determined hands-free, the aperture
photometry is straight-forward.  First, sky values are obtained as the median
flux in an annulus around each source, 30 to 60 pixels in radius; then, the
fluxes within circular apertures, less sky, are tabulated.

\begin{figure}
\plotfiddle{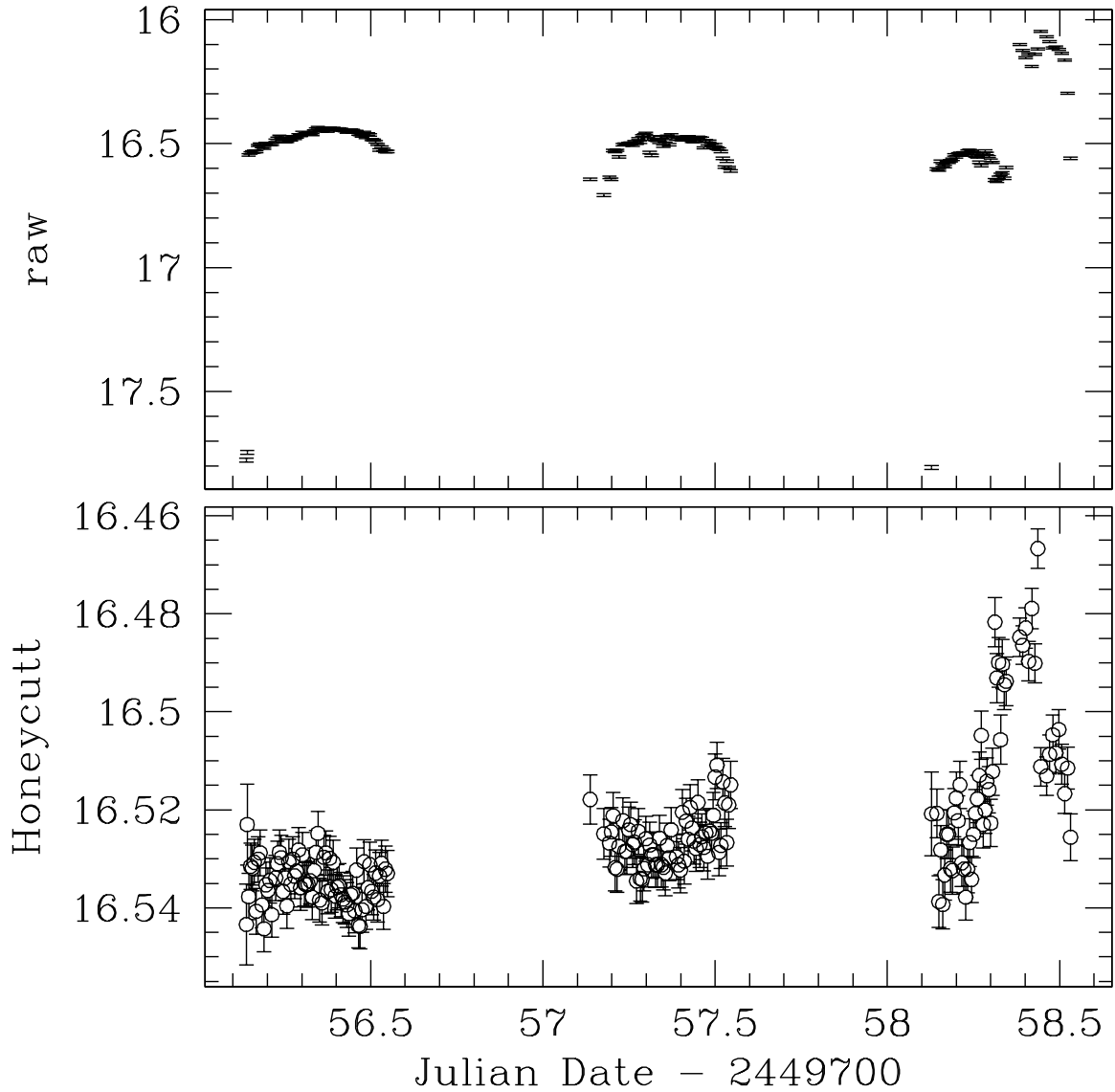}{14cm}{0}{80}{80}{-240}{-100}
\caption{Light-curve of ``A'' image of QSO 0957+561, before any correction
(top), and after the Honeycutt (1992) method for ensemble photometry is applied
(bottom).  Points have been omitted at top to allow the visibility of the
expected Poisson errorbars.  Errorbars at bottom also reflect Poisson noise
uncertainty.}
\label{figrawhon}
\end{figure}

At risk of being pedantic, we have plotted in Figure \ref{figrawhon} the raw
light-curve for the ``A'' (northern) image of Q0957+561 at top.  The errorbars
reflect Poisson uncertainty.  Because of weather and seeing, this raw aperture
photometry varies by more than 1.5 magnitudes, for an object that should have
approximately constant brightness over three nights.  With respect to the mean,
this light-curve yields a reduced $\chi^2$ of 1300 per degree of freedom.
Obviously, this is why one uses relative photometry (versus absolute
photometry), and at bottom, we have plotted the photometry of the A image
relative to that of comparison stars in the field, using the Honeycutt (1992)
method for ensemble photometry, and as expected the improvement is vast
($\chi^2 \sim 9$ per d.o.f.).

The Honeycutt (1992) method proves to be quite effective at producing a good
estimate of the relative brightness of objects on the same frame (c.f. Kundi\'c
\etal\ [1995 and 1997]), and this is the typical level of sophistication with
which light-curves of 0957+561 have been produced (ibid).  Such light-curves,
however, often contain nagging correlations between the ``A'' and ``B'' images
night to night.  After some effort to understand these correlations as possible
effects of moon phase, or other astronomical situations, the A-B correlations
remain largely unexplained.

Returning to Figure \ref{figrawhon}, we can see that even after the Honeycutt
correction, there remains a 5\% departure from the mean magnitude on the third
night.  This is particularly troubling for an object which is generally not
supposed to vary by more than about 0.01 magnitudes per day (Press \etal\
1992b; Keel 1982); hence, the 5\% variation over several hours must be due to
some observational condition.  We will show that this departure is due
principally to seeing, and its daughter effects, namely changing
characteristics of light from the galaxy, and cross contamination of light
between the two quasar images (separated by $6\arcsec$) during poor seeing.

\section{Galaxy Subtraction}

One effect of seeing arises because the giant elliptical lens galaxy, ``G1,''
has a rather cuspy profile.  This profile gets smeared over larger and larger
radii as seeing deteriorates.  Since the core of G1 is only $1\arcsec$ from the
B image, most of the galaxy's light lies within the image B aperture, but
outside the A aperture.  One would therefore expect poor seeing to introduce
more galaxy light into the image A aperture, and remove some light from the
image B aperture.  Since the galaxy has a an $R$ magnitude of 18.3 (about 15\%
of the image B brightness), changes of order 10\% in the galaxy's contribution
can introduce effects of order 1--2\% in the measured flux from images A and B,
a level that we are very much interested in.

Additionally, since 15\% is quite significant (the fraction would be smaller at
shorter wavelengths), removal of the galaxy is critical to proper treatment of
the cross-talk problem, which we will discuss in the next section---there we
will need to know the relative amounts of light in each aperture from the
quasar itself, before the apertures were polluted by the galaxy.

To correct for the galaxy contamination, we have taken advantage of published
HST imaging of galaxy G1 (Bernstein \etal\ 1997).  Those authors computed
best-fit elliptical isophotes for the galaxy, and we have used their plots of
ellipticity, position angle and surface magnitude (as functions of semi-major
axis, their Fig. 2.) to produce synthetic galaxy images for each data frame.

To determine the appropriate zero-point for the galaxy, we use the relative
magnitude shown in Bernstein \etal\ (1997) Fig. 2 between Kitt Peak {\it
R}-band data and {\it HST} \f555w data for the galaxy's brightness profile.  We
find that the correction, \f555w$-${\it R} = 1.2, is very close to the
correction expected for an elliptical galaxy at $z = 0.36$ (Fukugita,
Shimasaku \& Ichikawa 1995).  Note that no allowance for a color gradient has
been made.  Ideally, {\it HST} imaging in redder bands, such as {\it F606W}
and {\it F850W}, would exist, so that minimal color corrections would be
necessary.

In order to synthesize the galaxy, we start with an image 4 times oversampled
in both dimensions, and compute the semi-major axis at each pixel, while
carefully observing dependences of ellipticity, and position angle on
semi-major axis (Bernstein \etal\ 1997).  The pixels within this image are then
sampled dynamically, according to semi-major axis.  Those pixels closest to the
center are oversampled an additional $32 \times 32$ times, and those farther
out are sampled exponentially less frequently such that the farthest pixels are
sampled once.  In each subpixel is placed the correct flux, according to the
surface brightness dependence on semi-major axis (Bernstein \etal\ 1997).  The
dramatic oversampling ensures that the steeply increasing flux near the center
is well-sampled, and we find that this process results in flux error of a few
parts per 1000 in any given pixel (on the large $4 \times 4$ oversampled
image).

We then rotate and demagnify the $4 \times 4$ oversampled image, so that it
appears at the correct pixel scale and position angle according to astrometry
from the the data image.  The rotation and demagnification involves cubic
spline interpolation within the {{\it ROT}\/} routine provided by IDL.  We
convolve this properly sized and oriented image with the point spread function,
which we have previously obtained from comparison stars on the data frame.

To determine the proper zero-point for the galaxy, one must be aware that
simple aperture photometry from the comparison stars may not suffice.  When
seeing degrades, a non-negligible fraction of the light from a given star can
seep out of the aperture.  While the effect is similar for all point-sources on
the frame, the relative change in brightness differs significantly for resolved
objects, such as galaxy G1.  One should therefore calibrate a resolved object
with the ``total'' light from the comparison point-sources.  To assess the
``total'' light, we used a series of apertures of different size as input into
an implementation of {{\it Numerical Recipes' ratint.f}\/} rational function
extrapolator (Press \etal\ 1992a).  We could then
extrapolate to a very large aperture to estimate the total light.  However, in
practice, we found that even in the worst seeing conditions an aperture of
diameter 20 arcseconds contained all the light to the 1--2\% level, and avoided
the larger errors introduced by extrapolation.  We therefore calibrated the
synthesized galaxy flux with this very large aperture photometry from several
comparison stars on the field.

\begin{figure}
\plotfiddle{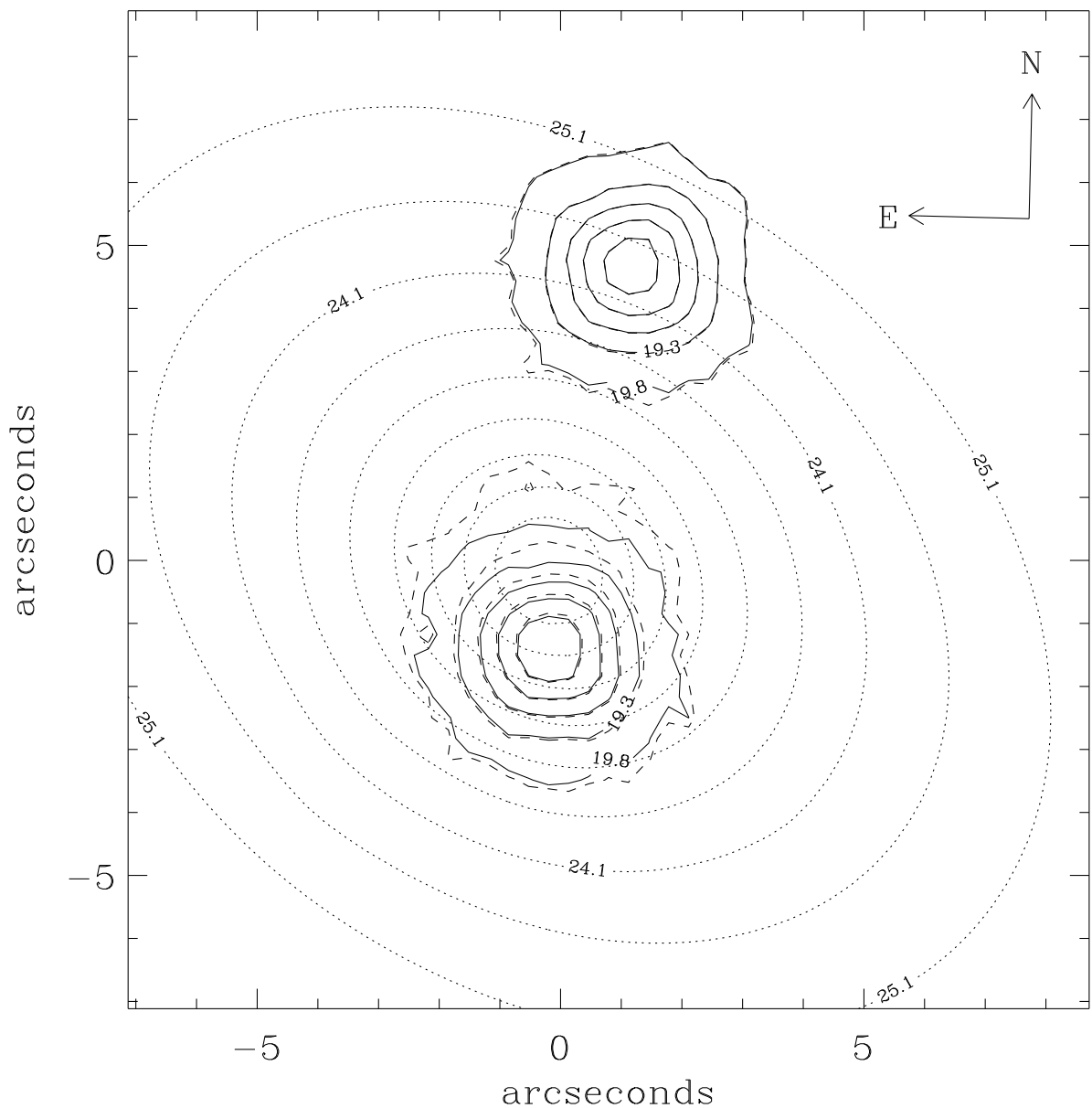}{14cm}{0}{70}{70}{-280}{-240}
\caption{Subtraction of the lens galaxy, G1, from images of the 0957+561
system.  The dotted contours are those of the galaxy image which is subtracted
from the data image.  The dashed contours represent the raw data image before
galaxy subtraction, with image A at top, and image B in bottom center.  The
solid contours are after galaxy subtraction.  Note the improved agreement in
shape of the quasar images after subtraction of the galaxy.  Please be aware
that the galaxy contour levels are much fainter than those of the quasar
images.}
\label{figgalsub}
\end{figure}

We now have a properly located, scaled and oriented convolved image to subtract
from the data frame, and after doing so, we perform the aperture photometry on
the A and B images once again.  In Figure \ref{figgalsub}, we have plotted a
contour map of the quasar images from a typical exposure, before and after the
galaxy subtraction.  The solid contours correspond to flux on the data frame
after galaxy subtraction, the dashed contours to flux before galaxy
subtraction.  We include, for reference, contours of the synthesized galaxy,
with much fainter contour levels, of course.  Notice the improvement in
similarity between the shapes of the image A and B contours: before galaxy
subtraction, the B image had significant extension in the northward direction,
due to the excess light from the galaxy.  The subtraction removes that
extension effectively.

In Figure \ref{figseegal}, we have plotted, as a function of seeing, the
fraction of total light in the A and B apertures due to the galaxy.  As
suspected, the image B photometry declines with poor seeing because more light
from the core of the galaxy is getting smeared out of the aperture.  The A
image, however, is far from the center of the galaxy, so that the smearing of
bright parts of the galaxy out to larger and larger radii increases the light
from the galaxy in the A image, which is visible in Figure \ref{figseegal}.
These two effects conspire to produce a relative change of up to 2.5\% from
good to bad seeing, which, in itself, should be corrected.

\section{Cross Talk Correction}

Another effect of seeing is ``cross talk'' between the images A and B,
particularly during bad seeing.  Since the images are separated by only
$6\arcsec$, apertures of $6\arcsec$ (diameter) practically rub against each
other, so light spilled out of one aperture due to seeing can very easily wind
up in the other aperture.

We have attempted to address this concern in a simple way.  After proper galaxy
(and sky) subtraction, the aperture fluxes of images A and B should represent
only light from the quasar.  If that's true, the comparison stars in the frame
should exhibit the same fraction of spillover of light into nearby apertures as
do the quasar images.

We selected apertures about nearby comparison stars; these apertures were
offset from the comparison stars at the same pixel offset as that between
images A and B (see Figure \ref{figseediag} test apertures AB and BA).  The
amount of light in these fiducial apertures as a fraction of the recorded
aperture flux from the comparison star should be identical to the fraction of
cross talk light from each quasar image.  We define $r_{BA}$ as the ratio of
the light in test aperture BA to the aperture flux of the comparison star, with
similar definition for $r_{AB}$.  Note that $r_{BA}$ estimates the ratio of the
spilled light from image B into the A aperture, to the actual light from image
A in the A aperture.

To correct for cross talk, we compute the corrected fluxes of A and B as
\begin{equation}
\label{eqncros}
\begin{array}{ll}
A_{corr} = ({A_{meas} - r_{BA}B_{meas}) \cdot
  (1-r_{AB}r_{BA}})^{-1}\\
B_{corr} = ({B_{meas} - r_{AB}A_{meas}) \cdot
  (1-r_{AB}r_{BA}})^{-1}\\
\end{array}
\end{equation}
where $A_{meas}$ is the measured flux, $A_{corr}$ is the corrected flux,
$r_{BA}$ is described above, and the complementary variables transposing A with
B obey the same notation.  The denominator compensates for the overcorrection
made while subtracting $r_{BA}B_{meas}$ when $B_{meas}$ is contaminated by
light from A.

We plot in Figure \ref{figseevsfrac} the $r_{AB}$ (triangles) and $r_{BA}$
(circles) as a function of FWHM seeing.  As expected, with good seeing, the
fraction is quite small, but with poor seeing, the fraction can approach 5\%,
which is a substantial correction for our purposes.

\section{Final Light-curves}

In Figures \ref{lightcurves} a, b and c, we have plotted the light-curves
from the nights February 6--8, 1995.  In each case we have used the Honeycutt
(1992) method for ensemble photometry to correct for the ``exposure magnitude''
as was done in Figure \ref{figrawhon}.

Figure \ref{lightcurves}a shows the raw aperture magnitudes of the A and B
images.  On the third night, particularly, there is an obvious hump in the
light-curve of both A and B, on the order of a few percent, but notably greater
in A than in B.  

\begin{figure}
\plotfiddle{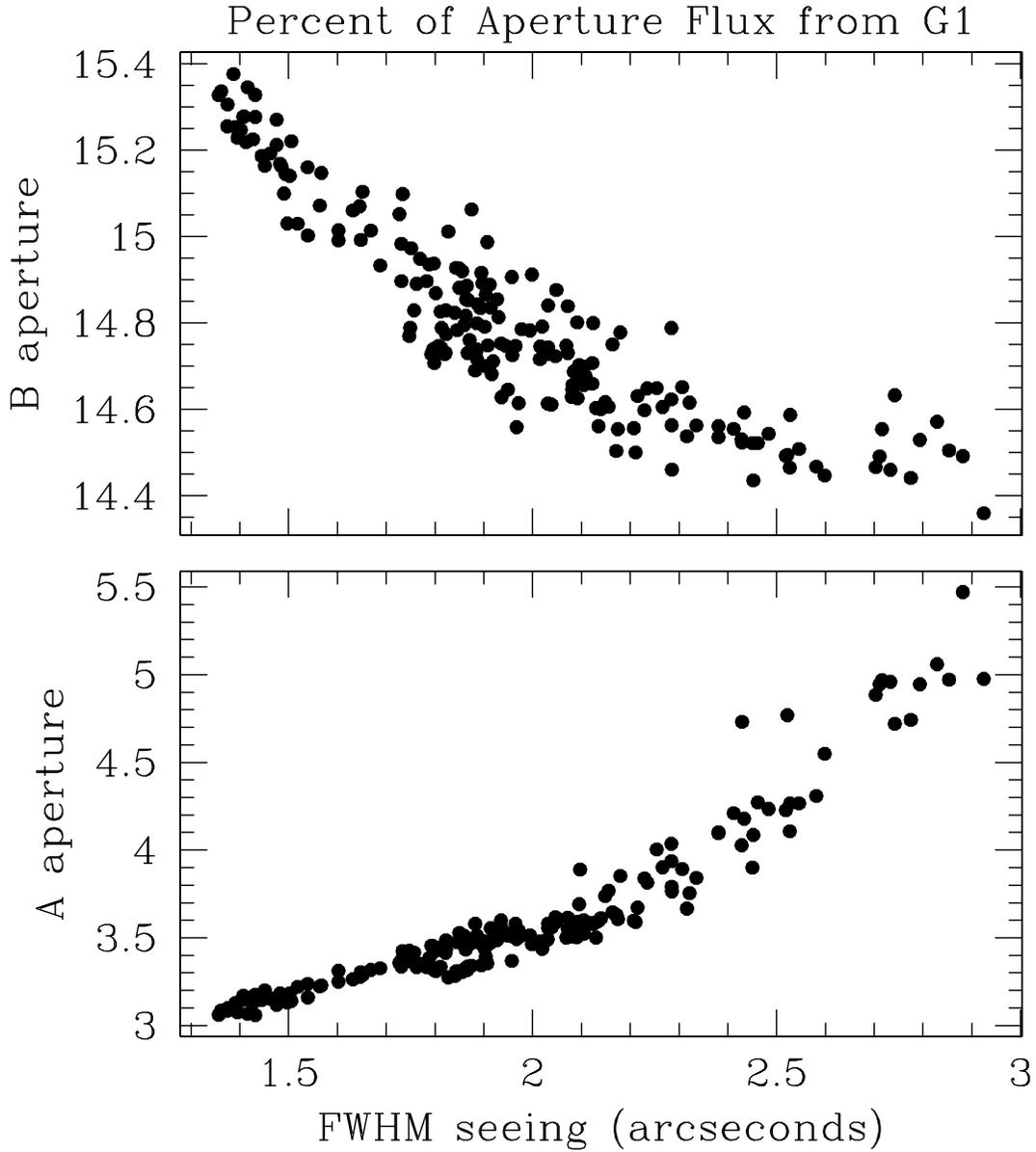}{14cm}{0}{80}{80}{-240}{-100}
\caption{The percentage contribution of light from the galaxy G1 to the
apertures about images A and B, as a function of seeing.  As seeing worsens the
galaxy's core is spread more thinly out of the B aperture, causing image B
apparently to dim, but image A brightens as seeing worsens, because more light
from the center of the galaxy is spread into its aperture.}
\label{figseegal}
\end{figure}

\begin{figure}
\plotfiddle{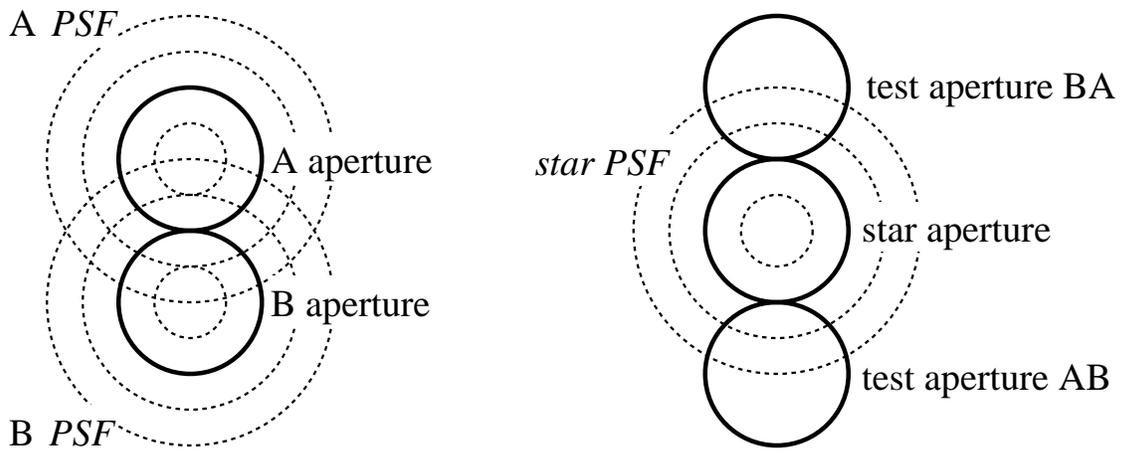}{14cm}{0}{150}{150}{-246}{-800}
\caption{The problem of cross talk.  At left, light from image A contaminates
the image B aperture, and vice versa, particularly in bad seeing.  At right,
the solution: test apertures are drawn near a comparison star at the same pixel
offsets exhibited by the A and B apertures.  The contamination of the test
apertures should be fractionally similar to the cross contamination of the A
and B apertures.}
\label{figseediag}
\end{figure}

\begin{figure}
\plotfiddle{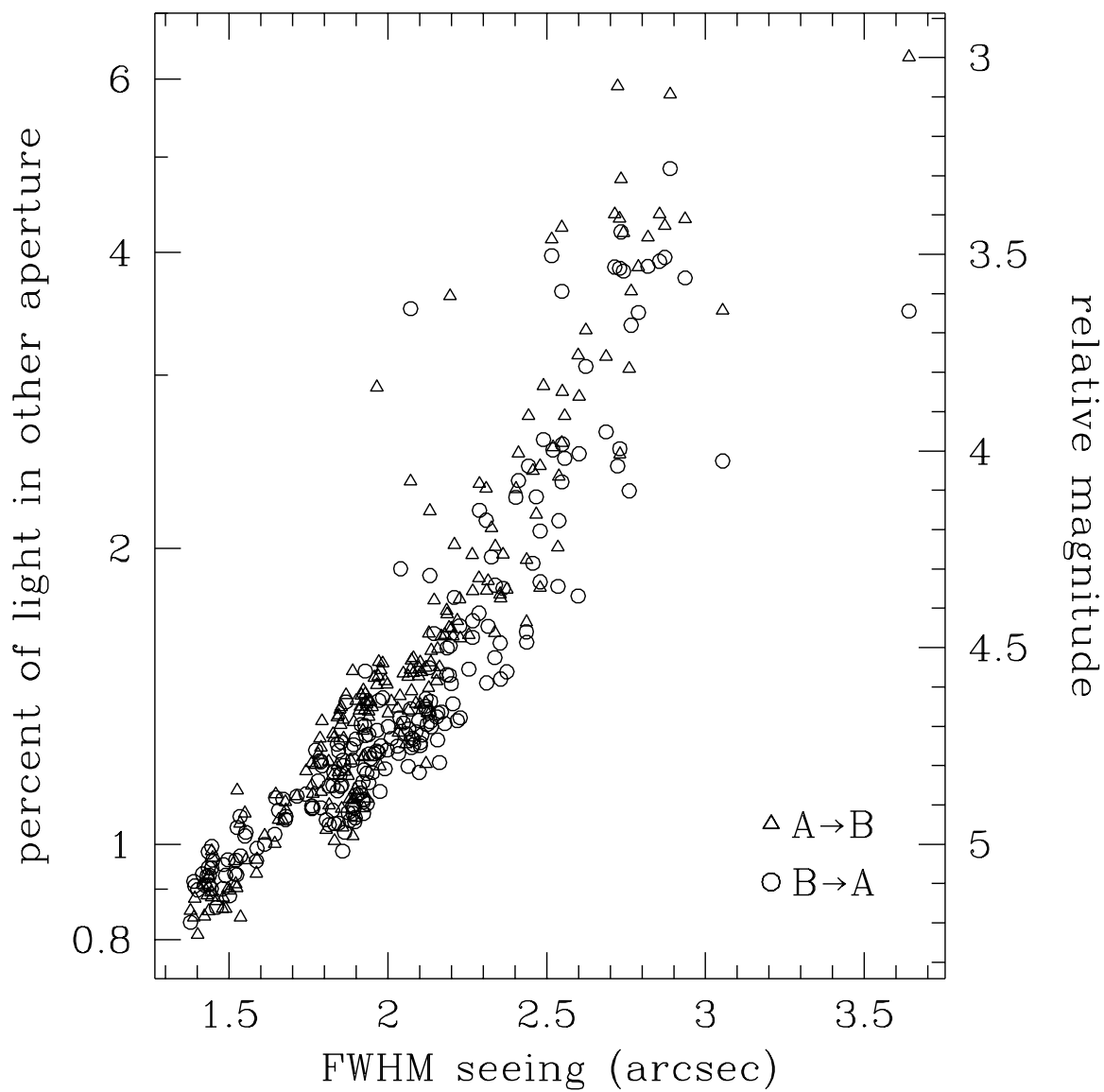}{14cm}{0}{80}{80}{-240}{-100}
\caption{Cross talk as a function of seeing.  The light from image A that
spills into aperture B (see Figure \ref{figseediag}), divided by the measured
aperture flux of image A, is plotted as triangles.  The corresponding
interchange of A and B is plotted as circles.  Notice that the ratios can
approach 5\%.}
\label{figseevsfrac}
\end{figure}

\begin{figure}
\plotfiddle{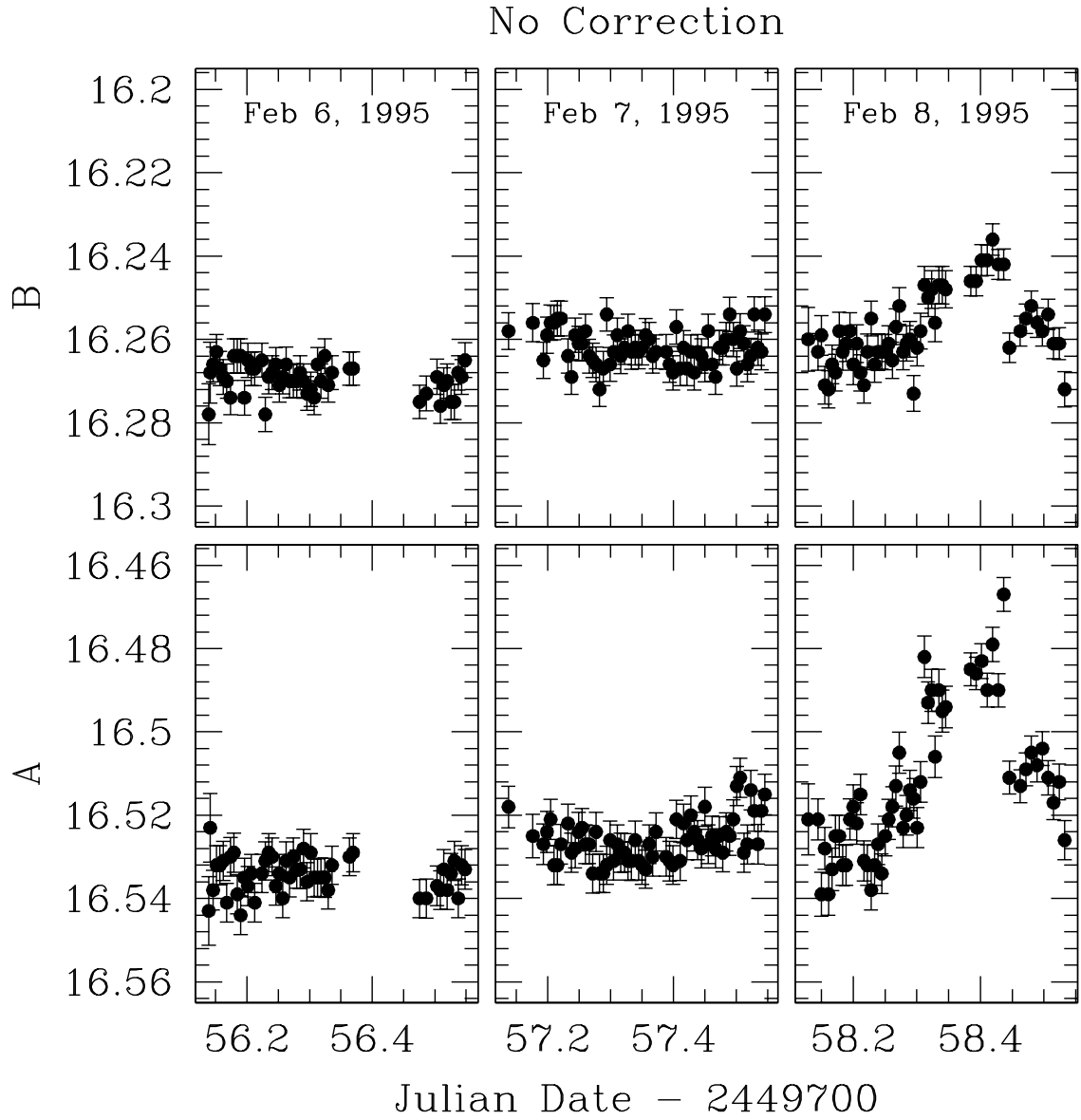}{14cm}{0}{80}{80}{-240}{-100}
\caption{a) The light-curve of images A and B after use of the Honeycutt (1992)
method for ensemble photometry.  Notice the large (few percent) hump on the
third night, which occurs in both the A and B images.}
\label{lightcurves}
\end{figure}

Figure \ref{lightcurves}b illustrates how the galaxy subtraction effects the
light-curves.  First of all, image B is dimmed by about 15\%, consistent with
Figure \ref{figseegal}.  The hump on the third night remains, but notice that
it has become more equal in amplitude in images A and B.  The hump itself has
occurred mainly because of the cross talk problem as the seeing degraded
substantially.  However, we can understand how the galaxy's light affects the
amplitude of the hump if we recall Figure \ref{figseegal}.  Figure
\ref{figseegal} demonstrated that as the seeing degraded, the galaxy
contributed more light to the A aperture, but less light to the B aperture.
Hence the seeing induced hump is more pronounced in A than in B.  Our galaxy
subtraction has allowed for this difference, and the amplitude of the hump has
come more into agreement between the two images. 

Figure \ref{lightcurves}c contains the final light-curves after correcting for
the cross talk problem.  Notice the removal of the hump on the third night.

The improvement in the light-curves after the two corrections can be
quantified, but we must first compute the errorbars.  In Figure
\ref{lightcurves}a, with no corrections, the errors are simply Poisson noise:
$\Delta \mbox{flux (in counts)} = [\mbox{flux + sky}]^{1/2}$.  In Figure
\ref{lightcurves}b, the total flux of the comparison stars for the galaxy has
been calculated, and added fractionally to the errobar:
\begin{equation}
\sigma^2_{A_{gal}} = A_{gal}^2 \cdot
\left[\;\sum_{i=1}^{N_*} \mbox{flux}_i\right]^{-1},
\end{equation}
where $A_{gal}$ is the flux in the A aperture from the galaxy, and the sum is
the total light from the comparison stars, which serve as the zeropoint for the
galaxy model.  This $\sigma_{A_{gal}}$ is, of course, added in quadrature to
the Poisson errorbars in Figure \ref{lightcurves}a.  Figures \ref{lightcurves}c
and d require an additional error term, from the comparison stars in the cross
talk calculation.  This term is rather more complicated, because a derivative
of equation \ref{eqncros} is required, so for sake of brevity, we will omit the
derivation.  Suffice it to say that the errorbar variances have the familiar
additional first order term,
\begin{equation}
\sigma^2(g) = \left({{\partial g}\over{\partial f}}\right)^2\sigma^2(f).
\end{equation}

In Table \ref{tab1}, we have listed the mean errorbar for each method.  As
expected they increase slightly with the introduction of the corrections.  We
have also included the (reduced) $\chi^2$ deviation from zeroth, first and
second order polynomial fits to the data, which can be seen in Figure
\ref{figfits} (one would not expect variations much more complicated than
second order, because the maximum change in one day is of order 1\% [Press
\etal\ 1992b]).  In each case the $\chi^2$ improves by at least 50\%, and as
much as 700\% from the raw data to the corrected data, with little penalty in
the mean error-bar of each point.  The corrected photometry is typically
reliable to about 0.55\% (5.5 millimagnitudes), within about 10\% ($\chi^2 \sim
1.2$) of Poisson error.  This can be compared with the roughly 5\% (50
millimagnitudes) reliability of the uncorrected A image (see Figure
\ref{lightcurves}a).

As a final comment, we note that Figure \ref{lightcurves} illustrates how a
zero time-delay correlation could appear in work such as Kundi\'c \etal\ (1995
and 1997).  The uncorrected light-curves of images A and B (Figure
\ref{lightcurves}a) appear very much correlated, as we can see in the last two
columns of Table \ref{tab1}, where we have listed the estimated Pearson
correlation coefficient ($\rho_{AB} = \sigma_{AB}\cdot[\sigma_A\sigma_B]^{-1}$)
for the data relative to the mean, and relative to the best-fit parabolas
($\rho_{AB,0}$ and $\rho_{AB,2}$, respectively).  Without getting into an
involved statistical discussion, we note that the value of of $\rho_{AB}$
improves dramatically after correction for cross talk.  The value of $\rho_{AB}
= 0.76$ has a chance of less than 1 in $10^{36}$ to arise from uncorrelated
Gaussian deviates, while a value of $\rho_{AB} = 0.22$ would arise 2.1 times
out of a thousand (Lupton, 1993).  After subtraction of the best-fit parabolas
for the three cases, the correlation coefficients $\rho_{AB,2}$ decrease in
each case, and while the value of 0.59 is only likely at the 1 in $10^{19}$
level, the value of 0.13 is 7\% likely to arise from uncorrelated Gaussian
variates.  While the exact meaning of these statistical probabilities is
debatable, the corrections obviously dramatically reduce statistical
correlations in the photometry.  These improvements make it easy to see that
seeing could explain some of the zero-delay correlations in Kundi\'c \etal\
(1995 and 1997) and in other monitoring projects as well.

Note, however, that the constancy of the quasar brightness found here is
comparable to that of photometry reported in Schild \& Thompson (1996).  This
is not surprising because the Schild photometry results from some censoring of
data during bad seeing (to avoid the problems detailed here).  Thus, for
example, data for Nov. 1998, show a total r.m.s. deviation of only 0.013
magnitudes, as reported in Schild (1990).

\section{Summary and Conclusions}

In order to search for microlensing in Q0957+561, it will be necessary to
process large number of data frames taken over more than a decade.  A highly
automated program for photometry is desirable.  We have therefore created a
hands-free system of generating reliable photometry for this object.

Furthermore, we have introduced methods to treat the problems related to
changing seeing that have plagued photometry of this object in the past.  One
problem is that seeing changes the amount of light introduced by the lens
galaxy into apertures about the A and B images.  A second problem is that bad
seeing causes cross contamination of light from one image into the other.
These effects conspire to cause correlated errors of up to 5\% in the
photometry.  This is at approximately the same level as variations expected
from microlensing by small stars and MACHO's in the lens galaxy, and is thus
necessary to remove.

We have greatly improved the situation by 1) using HST data to subtract the
lens galaxy, 2) using comparison stars to estimate the level of cross talk
between the images.  After these corrections, we find that the photometry is
reliable to about 5.5 millimagnitudes (0.55\%) for a single image frame, a vast
improvement over 5\% in the uncorrected photometry.  The fact that the
corrections for seeing practically eliminate correlated errors suggests that
the seeing is the principal cause for these errors.

With these methods to correct for seeing, we plan to reduce large amounts
of archival data on Q0957+561 and search for microlensing by discrete objects
within the lens galaxy.  To date, a handful of events at the 5\% level are
suggested by observation (Schild 1996), which Schmidt \& Wambsganss (1998) have
used to put important limits on allowed masses for MACHO's (assuming small QSO
source size).  With much improved photometry on the many more observations
available to us, we will be able to characterize with some confidence the
masses and densities of MACHO's in the Q0957 lens galaxy.

\newpage

\begin{table}
\begin{center}
{\bf TABLE 1}

\vspace{1 cm}

\begin{tabular} {lcccccccccccc}  \tableline \tableline
& \multicolumn{4}{c}A & & \multicolumn{4}{c}B \\
\tableline
method & $\bar{\sigma}$ & $\chi^2_0$ & $\chi^2_1$ & $\chi^2_2$ & ~ &
         $\bar{\sigma}$ & $\chi^2_0$ & $\chi^2_1$ & $\chi^2_2$ & ~ &
$\rho_{AB,0}$ & $\rho_{AB,2}$\\ 

no correction        & 0.0047 & 9.41 & 4.76 & 4.38 & &
                       0.0042 & 3.61 & 2.02 & 2.07 & & 0.79 & 0.66\\  
galaxy subtracted    & 0.0048 & 5.45 & 3.02 & 2.86 & &
                       0.0046 & 4.27 & 2.15 & 2.19 & & 0.76 & 0.59\\
cross talk corrected & 0.0049 & 1.23 & 1.23 & 1.19 & &
                       0.0047 & 1.27 & 1.20 & 1.02 & & 0.22 & 0.13\\
\tableline

\end{tabular}
\end{center}
\caption{Mean errorbars ($\bar{\sigma}$, in magnitudes), computed from Poisson
statistics alone; and reduced $\chi^2$ of fits to the mean magnitude
($\chi^2_0$), a line ($\chi^2_1$), and a parabola ($\chi^2_2$), over three
nights (see Figure \ref{figfits}).  The last two column columns are the Pearson
correlation coefficients of A with B, relative to their means ($\rho_{AB,0}$)
and to their best-fit parabolas ($\rho_{AB,2}$).  The first row is before any
correction for the galaxy or cross talk.  The second row is tabulated after the
galaxy has been subtracted.  The third row follows correction for cross talk.}
\label{tab1}
\end{table}
\newpage

\addtocounter{figure}{-1}
\begin{figure}
\plotfiddle{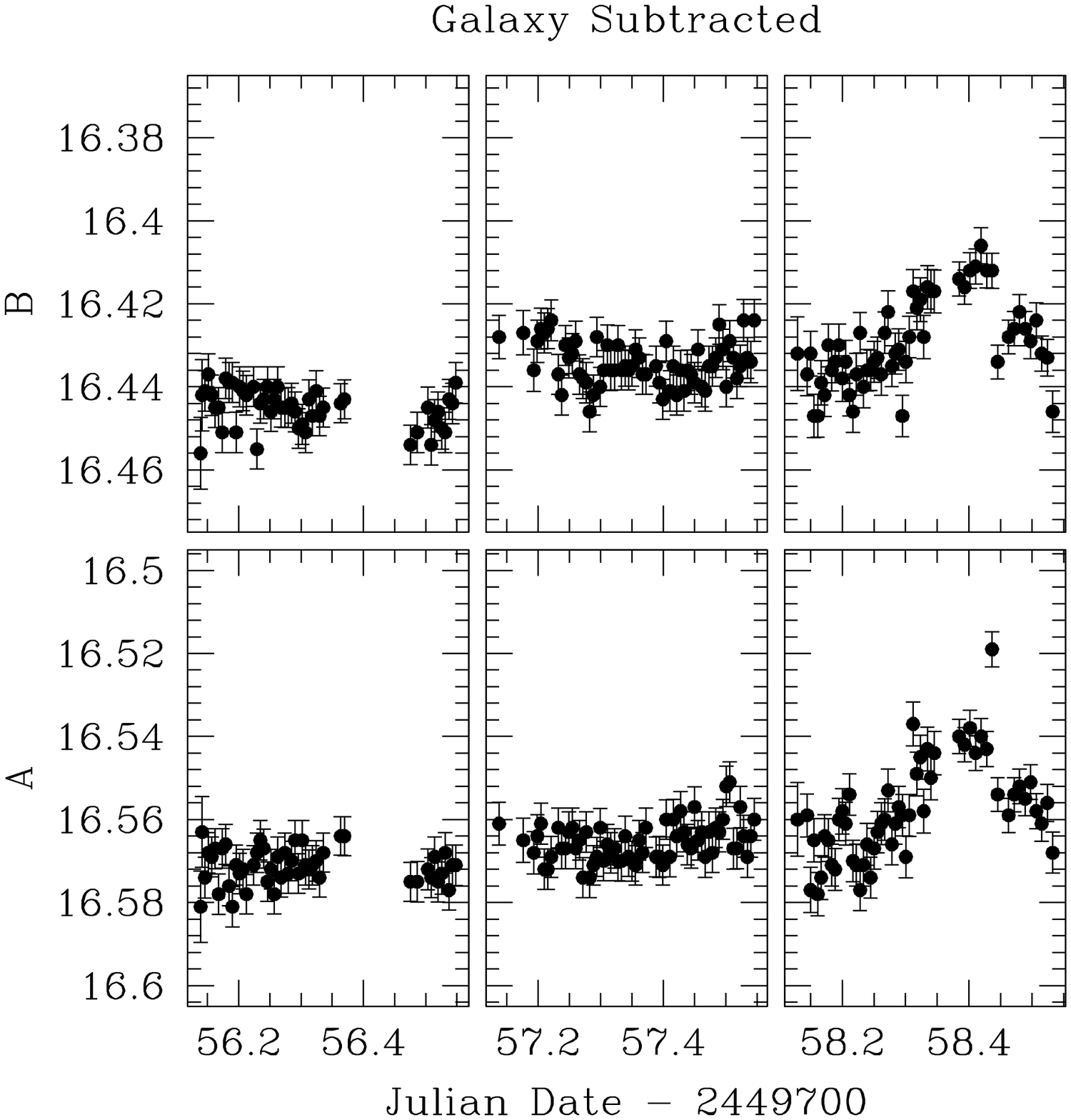}{14cm}{0}{80}{80}{-240}{-100}
\caption{b) The light-curve of images A and B after subtraction of the galaxy
G1's contribution to the A and B apertures.  The large hump in the A
light-curve has been reduced to a level similar to that of the B light-curve.}
\end{figure}

\addtocounter{figure}{-1}
\begin{figure}
\plotfiddle{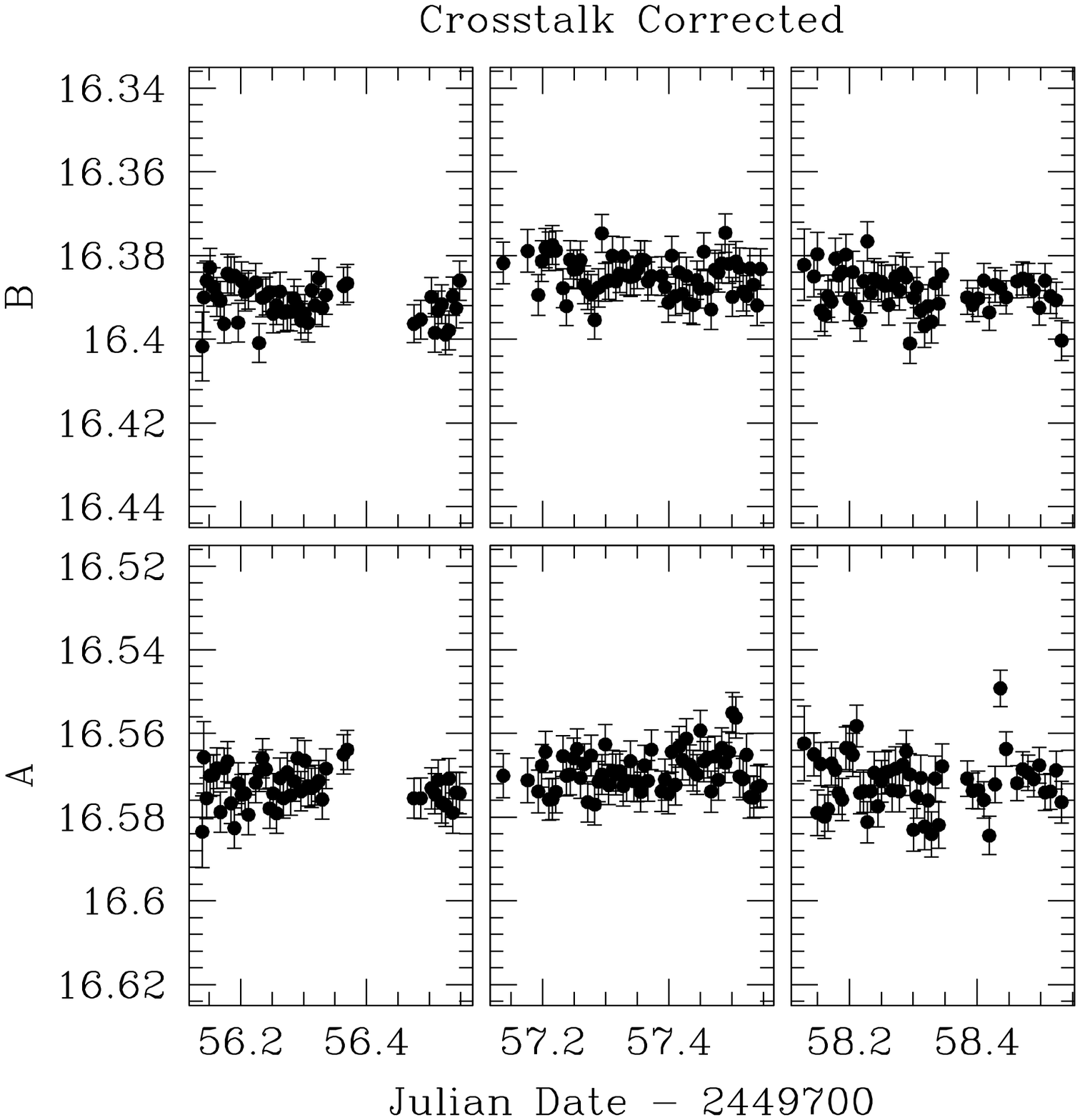}{14cm}{0}{80}{80}{-240}{-100}
\caption{c) The light-curve of images A and B after subtraction of the galaxy
G1 and correction for ``cross talk.''  Cross talk is the spill-over light from
the quasar images A and B into each other.  Notice the remarkable improvement
over figures \ref{lightcurves} a) and b) on the third night.  The ``hump'' has
virtually disappeared.}
\end{figure}

\begin{figure}
\plotfiddle{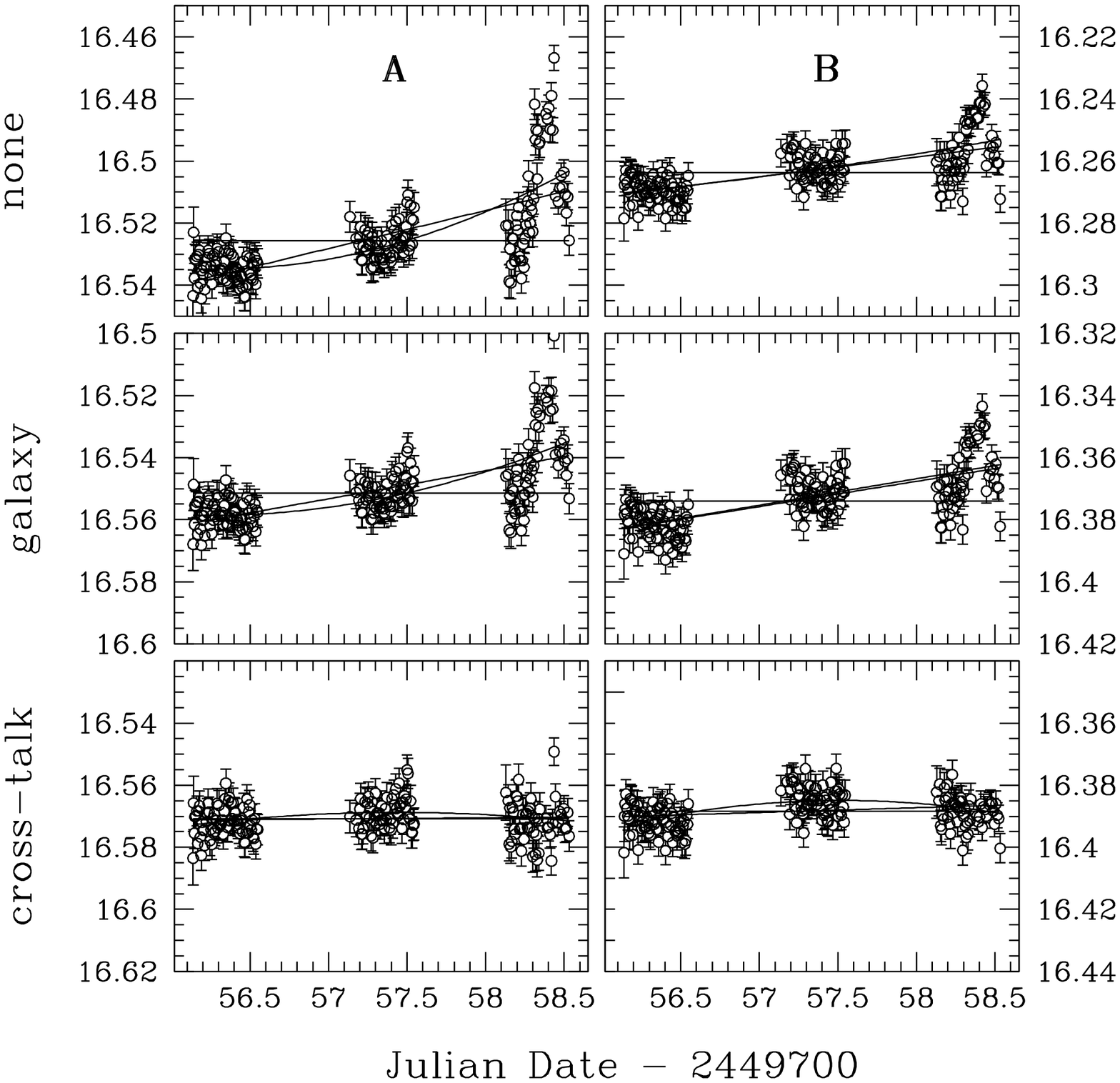}{14cm}{0}{80}{80}{-240}{-100}
\caption{The light-curves of images A (left) and B (right), before any
correction (top row), after subtraction of the galaxy G1 (middle row) and
correction for ``cross talk'' (bottom row).  Overplot are best fits for the
mean, a straight line, and for a parabola.}
\label{figfits}
\end{figure}

\end{document}